\documentclass[3p]{elsarticle}
\usepackage{lineno,hyperref}
\usepackage{amsfonts}
\usepackage{float}
\usepackage{amsmath}
\usepackage{amssymb}
\usepackage{hyperref}
\usepackage{color}
\usepackage{graphicx}
\usepackage{amssymb}
\usepackage{amsmath}
\usepackage{graphicx}
\usepackage{dcolumn}
\usepackage{bm}
\usepackage{epsfig}
\usepackage[T1]{fontenc}
\usepackage{ae,aecompl}
\usepackage{booktabs}
\usepackage{tabularx}
\usepackage{epsfig}
\usepackage{subfigure}
\modulolinenumbers[10]

\journal{Computers $\&$ Fluids}

\begin{document}
\begin{frontmatter}

\title{Cascaded lattice Boltzmann method for incompressible thermal flows with heat sources and general thermal boundary conditions}

\author[mymainaddress]{Linlin Fei}

\author[mymainaddress,mysecondaryaddress]{Kai Hong Luo\corref{mycorrespondingauthor}}
\cortext[mycorrespondingauthor]{Corresponding author}
\ead{K.Luo@ucl.ac.uk}

\address[mymainaddress]{ Center for Combustion Energy; Key laboratory for Thermal Science and Power Engineering of Ministry of Education, Department of Energy and Power Engineering, Tsinghua University, Beijing 100084, China}

\address[mysecondaryaddress]{Department of Mechanical Engineering, University College London, Torrington Place, London WC1E 7JE, UK}

\begin{abstract}
Cascaded or central-moment-based lattice Boltzmann method (CLBM) is a relatively recent development in the
LBM community, which has better numerical stability and naturally achieves better Galilean invariance for a specified lattice compared with the classical single-relation-time (SRT) LBM. Recently, CLBM has been extended to simulate thermal flows based on the double-distribution-function (DDF) approach [L. Fei \textit{et al.}, Int. J. Heat Mass Transfer 120, 624 (2018)]. In this work, CLBM is further extended to simulate thermal flows involving  complex thermal boundary conditions and/or a heat source. Particularly, a discrete source term in the central-moment space is proposed to include a heat source, and a general bounce-back scheme is employed to implement thermal boundary conditions. The numerical results for several canonical problems are in good agreement with the analytical solutions and/or numerical results in the literature, which verifies the present CLBM implementation for thermal flows. 
\end{abstract}

\begin{keyword}
CLBM \sep  thermal flows  \sep heat sources  \sep boundary conditions
\MSC[2010] 00-01\sep  99-00
\end{keyword}

\end{frontmatter}

\linenumbers

\section{Introduction}
In the last three decades or so, the lattice Boltzmann method (LBM), which is a mesoscopic numerical method based on the kinetic theory, has been developed to be a 
widely used numerical method for solving various fluid flows and heat transfer problems \cite{mcnamara1988use,higuera1989lattice,qian1992lattice,qian1995recent,chen1998lattice,gan2015discrete,lin2017multi}. In the LBM, a discretized Boltzmann equation, based on a specific discrete velocity set and designed to reproduce the Navier-Stokes (N-S) equations in the macroscopic limit, is solved for the distribution functions (DFs). Generally, the mesoscopic
nature of LBM allows its natural incorporation of microscopic and/or mesoscopic physical phenomena, while the highly efficient algorithm makes it affordable computationally \cite{succi2001lattice,li2016lattice}.

In the extensively used algorithm for LBM, the numerical process can be split into two steps \cite{succi2001lattice,guo2013lattice,li2016lattice}: the ``collision" step and the ``streaming" step. In the collision step, the single-relaxation-time (SRT) or BGK scheme \cite{qian1992lattice} is the most widely used  collision operator. In the BGK-LBM, all the distribution functions are relaxed to their local equilibrium states at an identical rate, where the relaxation rate is related to the kinematic viscosity. The BGK-LBM is quite simple to implement and can simulate low Reynolds number flows, but it may have numerical instability at high Reynolds number or low-viscosity flows, as well as inaccuracy of implementing the boundary conditions \cite{d1994generalized,lallemand2000theory,ginzburg2003multireflection,pan2006evaluation,guo2008analysis}. To overcome these difficulties, the multiple-relaxation-time (MRT) collision operator was proposed in the literature \cite{d1994generalized,lallemand2000theory}. In the MRT-LBM, the DF is transformed into a raw moment space, where different raw moments of the DF can be relaxed at different relaxation rates to the local equilibrium raw moments. Compared with the BGK-LBM, the MRT-LBM can enhance numerical stability by carefully separating the time scales among the kinetic modes \cite{lallemand2000theory}, as well as improve the numerical accuracy for non-slip boundary conditions by choosing a specified relaxation rate for the energy flux \cite{ginzburg2003multireflection,pan2006evaluation,guo2008analysis}.
However, Geier \textit{et al.} argued that the MRT-LBM may also encounter instability for high Reynolds number flows due to the insufficient degree of Galilean invariance and the ``cross-talk" effect induced by relaxing the raw moments \cite{geier2006cascaded}. By relaxing central moments of the DF in the co-moving frame, a cascaded or central-moment-based operator was proposed in 2006 \cite{geier2006cascaded}. In the cascaded LBM, also known as CLBM, the ``cross-talk" effect in the MRT-LBM is eliminated naturally, and a higher degree of Galilean invariance for a specified lattice can be preserved readily by matching the higher order central moments of the continuous
Maxwell-Boltzmann distribution. By setting the relaxation rates for high-order central moments to be 1, CLBM has been applied to simulate high Reynolds number ($ Re=1400000 $) turbulent flow using coarse grids without resorting to any turbulence models \cite{geier2006cascaded}. Recently, CLBM has been extended to simulate multiphase flows coupled with the pseudo-potential model \cite{shan1993lattice} by Lycett-Brown and Luo \cite{lycett2014multiphase}. Compared with the BGK-LBM for multiphase flows, the proposed multiphase CLBM 
reduces the spurious currents near the phase interface significantly \cite{lycett2014multiphase}, and achieves higher stability range for the Reynolds number \cite{lycett2014binary}.
As is known, the basic pseudo-potential model has some drawbacks, such as thermodynamic inconsistency, large spurious currents, and suffers from the problem of tuning the surface tension independently of the density ratio \cite{li2016lattice}. More recently,
Li \textit{et al.} proposed an approach of achieving thermodynamic consistency via tuning the mechanical stability condition \cite{li2012forcing,li2013lattice}, and analyzed the effects of the equation of state on the thermodynamic consistency \cite{li2014thermodynamic}.
Inspired by the methods in  \cite{li2012forcing,li2013lattice,li2014thermodynamic}, an improved forcing scheme in pseudo-potential model was proposed in \cite{lycett2015improved}. By coupling the improved forcing scheme with the cascaded operator, Lycett-Brown and Luo 
achieved very high parameters in the simulation of binary droplet collision\cite{lycett2016cascaded}.

More recently, CLBM was first extended to simulate thermal flows by the present authors \cite{fei2016thermal}, where a thermal cascaded lattice Boltzmann method (TCLBM) was proposed based on the double-distribution-function (DDF) approach. In our TCLBM, the CLBM is used to simulate the flow field and another total energy BGK-LBM is used for the temperature field, where the two fields are coupled by equation of state for the ideal gas. The proposed TCLBM has been proved to be able to simulate low-Mach compressible thermal flows with commendable stability and accuracy. For incompressible thermal flows without viscous dissipation and pressure work, another CLBM has been constructed on a simpler lattice (D2Q5) to solve the passive-scalar temperature field \cite{fei2018modeling}. Compared with the D2Q5 MRT-LBM for the temperature equation, the proposed D2Q5 CLBM is shown to be better Galilean invariant. Thus a higher characteristic velocity can be adopted for convection heat transfer problems, which decreases the computational load significantly. Although CLBM has been applied to several thermal problems \cite{fei2016thermal,fei2018modeling}, less attention has been paid to two important factors: temperature field with a heat source and non-isothermal boundary conditions. In this work, we will present the implementation of a heat source and a general bounce-back scheme for the thermal boundary conditions.

The rest of the paper is structured as follows: In Section \ref{sec.2},  a brief introduction for the DDF-based CLBM for incompressible thermal flows is given, followed by the implementation of a heat source and general bounce-back scheme for thermal boundary conditions. Numerical experiments are carried out for several benchmark problems to validate the employed method in  Section \ref{sec.3}. Finally, a brief summary is given in Section \ref{sec.4}.

\section{Numerical method}\label{sec.2}
The macroscopic governing equations for incompressible thermal flows can be written as:
\begin{subequations}\label{e1}
	\begin{equation}\label{e1a}
\nabla  \cdot {\bf{u}} = 0,
	\end{equation} 
	\begin{equation}\label{e1b}
\frac{{\partial {\bf{u}}}}{{\partial t}} + {\bf{u}} \cdot \nabla {\bf{u}} =  - \frac{1}{{{\rho _0}}}\nabla p + \nu {\nabla ^2}{\bf{u}} + {\bf{F}},
	\end{equation}
	\begin{equation}\label{e1c}
	\frac{{\partial T }}{{\partial t}} + {\bf{u}} \cdot \nabla T  = \nabla  \cdot (\alpha \nabla \phi ).
	\end{equation} 
\end{subequations}
where $ {\bf{u}}  $, $ {p} $, $ {\rho _0} $, $ T $, $ \nu $ and $ \alpha $ are the velocity, pressure, reference density, temperature, kinematic viscosity, and thermal diffusivity, respectively. The Boussinesq approximation is employed in this work, thus the force field is defined as,
\begin{equation}\label{e2}
{\bf{F}} =  - {\bf{g}}\beta (T - {T_0}) + {{\bf{F}}_v},
\end{equation}
where the gravitational acceleration vector $ {\bf{g}} $ points to the  negative direction of y-axis, $ \beta $ is the thermal expansion coefficient, $ {T_0} $ is the reference temperature, and $ {\bf{F}}_v $ is an external body force.

\subsection{CLBM for the flow field}\label{sec.2.1}
In the present work, the D2Q9 discrete velocity model \cite{qian1992lattice} is used to simulate two-dimensional problems. As usual,  the lattice spacing $ \Delta{x}$, time step $ \Delta {t} $ and lattice speed  $ c=\Delta{x}/\Delta {t} $ are set to be 1. The discrete velocities ${{\bf{e}}_i} = \left[ {\left| {{e_{ix}}} \right\rangle ,\left| {{e_{iy}}} \right\rangle } \right]
$ are defined by
\begin{subequations}\label{e3}
	\begin{equation}
	\left| {{e_{ix}}} \right\rangle  = {[0,1,0, - 1,0,1, - 1, - 1,1]^\top}, \\ 
	\end{equation} 
	\begin{equation}
	\left| {{e_{iy}}} \right\rangle  = {[0,0,1,0, - 1,1,1, - 1, - 1]^\top}, \\ 
	\end{equation} 
\end{subequations}
where $ i = 0...8 $, ${\left|  \cdot  \right\rangle }$ denotes the column vector, and the superscript $ \top $ indicates transposition.

For the cascaded collision operator, the collision step is carried out in the central-moment space. The raw moments and central moments of the discrete distribution functions (DFs) ${{f_i}}$ are defined as:
\begin{subequations}\label{e4}
	\begin{equation}
	{k_{{mn}}} = \left\langle {{f_i}\left| {e_{ix}^me_{iy}^n} \right.} \right\rangle , \\  
	\end{equation} 
	\begin{equation}
	{{\tilde k}_{{mn}}} = \left\langle {{f_i}\left| {{{({e_{ix}} - {u_x})}^m}{{({e_{iy}} - {u_y})}^n}} \right.} \right\rangle , \\
	\end{equation} 
\end{subequations}
and the equilibrium values $ k_{_{{mn}}}^{eq} $ and $ \tilde k_{{mn}}^{eq}$ are defined
analogously by replacing ${{f_i}}$ with the discrete equilibrium distribution functions (EDFs) $ {f_i^{eq}} $. In this work, a simplified raw-moment set is adopted \cite{fei2018modeling},
\begin{equation}\label{e5}
\left| {{\Gamma_i}} \right\rangle  = 
\left[ {{k_{00}},{k_{10}},{k_{01}},{{k_{20}},{k_{02}}},{k_{11}},{k_{21}},{k_{12}},{k_{22}}} \right]^\top,
\end{equation}
and so do the central moments $ \tilde \Gamma_i$. Specifically, the raw moments can be given from ${{f_i}}$  through a transformation matrix $ {\bf{M}}$ by $ \left| {{\Gamma _i}} \right\rangle  = {\bf{M}}\left| {{f_i}} \right\rangle $, and the central moments shifted from raw moments can be performed through a shift matrix $ {\bf{N}} $ by $ \left| {{{\tilde \Gamma }_i}} \right\rangle  = {\bf{N}}\left| {{\Gamma _i}} \right\rangle $. The formulations for $ {\bf{M}}$ and $ {\bf{N}} $ can be easily 
obtained according to the raw-moments set \cite{fei2017consistent}. In the present study, $ {\bf{M}}$ and $ {\bf{N}} $ are expressed as \cite{fei2018modeling},
\begin{subequations}\label{e6}
	\begin{equation}
	{\bf{M}} = \left[ 
	\begin{array}{c c c c c c c c c}
	1 &1 &1&1&1&1&1&1&1\\
	0&1&0&-1&0&1&-1&-1&1\\
	0&0&1&0&-1&1&1&-1&-1\\
	0&1&0&1&0&1&1&1&1\\
	0&0&1&0&1&1&1&1&1\\
	0&0&0&0&0&1&-1&1&-1\\
	0&0&0&0&0&1&1&-1&-1\\
	0&0&0&0&0&1&-1&-1&1\\
	0&0&0&0&0&1&1&1&1\\
	\end{array} 
	\right], 
	\end{equation} 
	\begin{equation}
	{\bf{N}} = \left[ 
	\begin{array}{c c c c c c c c c}
	1 &0 &0&0&0&0&0&0&0\\
	-{u_x}&1&0&0&0&0&0&0&0\\
	-{u_y}&0&1&0&0&0&0&0&0\\
	u_x^2&-2{u_x}&0&1&0&0&0&0&0\\
	u_y^2&0&-2{u_y}&0&1&0&0&0&0\\
	{u_x}{u_y}&-u_y&-u_x&0&0&1&0&0&0\\
	-u_x^2{u_y}&2{u_x}{u_y}&u_x^2&- {u_y}	
	&0&-2u_x&1&0&0\\
	-u_y^2{u_x}&{u_y}^2&2{u_x}{u_y}&0	
	&-{u_x}&-2u_y&0&1&0\\
	u_x^2u_y^2&-2{u_x}u_y^2&-2{u_y}u_x^2&u_y^2&u_x^2&4{u_x}{u_y}&-2{u_y}&-2{u_x}&1\\
	\end{array} 
	\right].
	\end{equation} 
\end{subequations}
The post-collision central moments can be obtained by relaxing each of them to their
local equilibrium states independently,
\begin{equation}\label{e7}
\begin{aligned}
\left| {\tilde \Gamma_i^*} \right\rangle = ({\bf{I - S}})\left| {{{\tilde \Gamma}_i}} \right\rangle  + {\bf{S}}\left| {\tilde \Gamma_i^{eq}} \right\rangle  + ({\bf{I - S}}/2)\left| {{C_i}} \right\rangle, \\   
\end{aligned}
\end{equation}
where the block-diagonal relation matrix is given by,
\begin{equation}\label{e8}
{\bf{S}} = diag\left( {[0,0,0],\left[ \begin{array}{l}
	{s_ + },{s_ - } \\ 
	{s_ - },{s_ + } \\ 
	\end{array} \right],[{s_v},{s_3},{s_3},{s_4}]} \right),
\end{equation}
with ${s_ + } = ({s_b} + {s_\nu})/2$ and ${s_ - } = ({s_b} - {s_\nu})/2$ \cite{Asinari2008Generalized,fei2018modeling}. The kinematic $ \nu $ and bulk viscosities $ \nu_b $ are related to the relaxation parameters by $ \nu=  (1/s_\nu-0.5)/3 $ and  $ \nu_b=  (1/s_b-0.5)/3 $, respectively.

The equilibrium central moments $ {\tilde \Gamma_i^{eq}} $ are defined equal to the continuous central moments of the Maxwellian-Boltzmann distribution in continuous velocity space,
\begin{equation}\label{e9}
\left| {\tilde \Gamma_i^{eq}} \right\rangle  = \left[ {\rho ,0,0,\rho c_s^2,\rho c_s^2,0,0,0,\rho c_s^4} \right]^\top,
\end{equation}
where $ \rho $ is the fluid density, and $ c_s=\sqrt{1/3} $ is the lattice sound speed. Consistently, the forcing source terms in central moments space are given by \cite{fei2017consistent},
\begin{equation}\label{e10}
\left| {{C_i}} \right\rangle  = {[0,{F_x},{F_y},0,0,0,c_s^2{F_y},c_s^2{F_x},0]^ \top }.
\end{equation}
It may be noted that the method of incorporating a force field into the CLBM is the most recently proposed consistent forcing scheme \cite{fei2017consistent} and it shows great advantages over the previous forcing schemes in CLBM.  

In the  streaming step, the post-collision discrete DFs $ f_i^* $ in space $ \bf{x} $ and time $ t $ stream to their neighbors in the next time step as usual,
\begin{equation}\label{e11}
{f_i}(\textbf{x} + {\textbf{e}_i}\Delta t,t + \Delta t) = f_i^*(\textbf{x},t),
\end{equation}
where the post-collision discrete DFs are determined by $ \left| {f_i^*} \right\rangle  = {{\bf{M}}^{ - 1}}{{\bf{N}}^{ - 1}}\left| {\tilde \Gamma_i^*} \right\rangle$.
Using the Chapman-Enskog analysis, the incompressible N-S equaltions in Eqs. (\ref{e1}) can be reproduced in the low-Mach number limit \cite{fei2017consistent,premnath2009incorporating}. The hydrodynamics variables are obtained by,
\begin{equation}\label{e12}
\rho  = \sum\nolimits_i {{f_i}} ,~~~\rho {\bf{u}} = \sum\nolimits_i {{f_i}} {{\bf{e}}_i} + \frac{{\Delta t}}{2}{\bf{F}}.
\end{equation}
It should be noted that the incompressible approximation \cite{he1997lattice} is employed in the present work. Thus the dynamic variable density $ \rho $ can be divided into the reference density $ {\rho _0} $ and a small density fluctuation $ \delta \rho $.

\subsection{CLBM for the temperature field}\label{sec.2.2}
Due to the simplicity of convection-diffusion equation, a D2Q5 discrete velocity model (the five discrete velocity set is defined in Eq. (\ref{e3}), $ \left\{ {{{\bf{e}}_i} = \left[ {\left| {{e_{ix}}} \right\rangle ,\left| {{e_{iy}}} \right\rangle } \right]\left| {i = 0,1,...4} \right.} \right\}
 $) can be used to construct the CLBM for the temperature field \cite{fei2018modeling}. Similarly, the raw moments and central moments of the temperature distribution functions $ {g_i} $ are  defined by \cite{fei2018modeling}, 
\begin{subequations}\label{e13}
	\begin{equation}
	{k_{mn}^T = \left\langle {{g_i}\left| {e_{ix}^me_{iy}^n} \right.} \right\rangle ,} 
	\end{equation} 
	\begin{equation}
	 {\tilde k_{mn}^T = \left\langle {{g_i}\left| {{{({e_{ix}} - {u_x})}^m}{{({e_{iy}} - {u_y})}^n}} \right.} \right\rangle}.
	\end{equation} 
\end{subequations}
In the D2Q5 lattice, the following five raw moments are adopted \cite{fei2018modeling},
\begin{equation}\label{e14}
\left| {\Gamma _i^T} \right\rangle  = {\left[ {k_{00}^T,k_{10}^T,k_{01}^T,k_{20}^T, k_{02}^T} \right]^ \top },
\end{equation}
and so do the central moments $ \left| {\tilde \Gamma _i^T} \right\rangle $.
Analogously, the raw moments and central moments can be calculated through a transformation matrix $ {{{\bf{M}}_\textbf{T}}} $ and a shift matrix $ {{{\bf{N}}_\textbf{T}}}$, respectively \cite{fei2018modeling},
\begin{equation}\label{e15}
{\left| {\Gamma _i^T} \right\rangle  = {{\bf{M}}_\textbf{T}}\left| {{g_i}} \right\rangle,}~~~
{\left| {\tilde \Gamma _i^T} \right\rangle  = {{\bf{N}}_\textbf{T}}\left| {\Gamma _i^T} \right\rangle .}
\end{equation} 
Explicitly, the transformation matrix $ {{{\bf{M}}_\textbf{T}}} $ is expressed as \cite{fei2018modeling},
\begin{equation}\label{e16}
{{\bf{M}}_\textbf{T}}= \left[ 
\begin{array}{c c c c c}
1 &1 &1&1&1\\
0&1&0&-1&0\\
0&0&1&0&-1\\
0&1&0&1&0\\
0&0&1&0&1\\
\end{array} 
\right],
\end{equation}
and the shift matrix $ {{\bf{N}}_\textbf{T}} $ is given by,
\begin{equation}\label{e17}
{{\bf{N}}_\textbf{T}} = \left[ 
\begin{array}{c c c c c}
1 &0 &0&0&0\\
-{u_x}&1&0&0&0\\
-{u_y}&0&1&0&0\\
u_x^2&-2{u_x}&0&1&0\\
u_y^2&0&-2{u_y}&0&1\\
\end{array} 
\right].
\end{equation} 

The collision in the central-moment space can be written as,
\begin{equation}\label{e18}
\left| {\tilde \Gamma _i^{T,*}} \right\rangle  = ({\bf{I}} - {{\bf{S}}_\textbf{T}})\left| {\tilde \Gamma _i^T} \right\rangle  + {{\bf{S}}_\textbf{T}}\left| {\tilde \Gamma _i^{T,eq}} \right\rangle, 
\end{equation} 
where ${{\bf{S}}_\textbf{T}} = diag({\lambda _o},{\lambda _1},{\lambda _1},{\lambda _2},{\lambda _2})$ is the diagonal relaxation matrix. The thermal diffusivity is related to the relaxation parameter by  $\alpha  = (1/{\lambda _1} - 0.5)c_{T}^2\Delta t$. The equilibrium values of the central moments are given by,
\begin{equation}\label{e19}
\left| {\tilde \Gamma _i^{T,eq}} \right\rangle  = {\left[ {T,0,0,Tc_{T}^2,Tc_{T}^2} \right]^ \top },
\end{equation} 
where $ c_{T}^2 $ is the sound speed in the D2Q5 lattice. The post-collision temperature distribution functions $ g_i^*$ can be obtained by
\begin{equation}\label{e20}
g_i^* = {\bf{M}}_\textbf{T}^{ - 1}{\bf{N}}_\textbf{T}^{ - 1}\left| {\tilde \Gamma _i^{T,*}} \right\rangle. 
\end{equation} 
The streaming step for $ g_i^*$ is also as usual,
\begin{equation}\label{e21}
{g_i}({\bf{x}} + {{\bf{e}}_i}\Delta t,t + \Delta t) = g_i^*({\bf{x}},t). 
\end{equation}
The temperature $ T $ is computed as,
\begin{equation}\label{e22}
T = \sum\nolimits_{i = 0}^4 {{g_i}}.
\end{equation}
Through the Chapman-Enskog analysis, the convection-diffusion equation for the temperature field can be recovered in the macroscopic limit.
\subsection{Heat source and boundary conditions}\label{sec.2.3}
The DDF-based CLBM introduced above has been proved to be able to simulate several incompressible thermal flows with isothermal boundary condition. However, it can hardly simulate convective heat transfer problems with a heat source. Inspired by the previous method to include the heat source in the BGK and MRT LBM \cite{wang2007lattice,cui2016discrete}, here we present a CLBM for the temperature equation with a generalized heat source term. Similar to the consistent forcing scheme in CLBM, a heat source $ Q $ can be incorporated into Eq. (\ref{e18}) by means of central moments,
\begin{equation}\label{e23}
\left| {\tilde \Gamma _i^{T,*}} \right\rangle  = ({\bf{I}} - {{\bf{S}}_{\bf{T}}})\left| {\tilde \Gamma _i^T} \right\rangle  + {{\bf{S}}_{\bf{T}}}\left| {\tilde \Gamma _i^{T,eq}} \right\rangle  + ({\bf{I}} - {{\bf{S}}_{\bf{T}}}/2)\left| {{R_i}} \right\rangle.
\end{equation}
where $ {R_i} $ correspond to the central moments of the heat source,
\begin{equation}\label{e24}
\left| {{R_i}} \right\rangle  = \left[ {Q,0,0,Qc_{T}^2,Qc_{T}^2} \right].
\end{equation}
Analogously, the calculation of temperature is modified,
\begin{equation}\label{e25}
T = \sum\nolimits_{i = 0}^4 {{g_i}}  + Q/2.
\end{equation}

To implement thermal boundary conditions, a general bounce-back scheme is adopted in this work. After the collision step, the post-collision temperature distribution functions are obtained by Eq. (\ref{e20}). In the streaming step, the distribution functions entering from ``outside" of the boundary  ${g_{\vec i}}({{\bf{x}}_f},t + \Delta t)$ are determined by,
\begin{eqnarray}\label{e26}
{g_{\vec i}}({{\bf{x}}_f},t + \Delta t) =  - g_i^*({{\bf{x}}_f},t) + c_{T}^2{T_w},
\end{eqnarray}
where $ {{\bf{e}}_{\vec i}} =  - {{\bf{e}}_i} $, and ${T_w}$ is the temperature at the boundary. For the general thermal boundary conditions, ${b_1}{{\partial {T_w}} \mathord{\left/{\vphantom {{\partial {T_w}} {\partial n}}} \right. \kern-\nulldelimiterspace} {\partial n}} + {b_2}{T_w} = {b_3}$, the boundary temperature ${T_w}$ can be solved using a finite-difference scheme. Different from the 
method in \cite{zhang2012general}, a second-order finite-difference scheme is adopted for the temperature gradient,
\begin{equation}\label{e27}
\frac{{\partial {T_w}}}{{\partial n}} = \frac{{8{T_w} - 9{T_1} + {T_2}}}{{ 3{\bf{n}} \cdot {{\bf{e}}_i}\Delta x}},
\end{equation}
where $ T_1 $ and $ T_2 $ are temperatures at the first and second layer nodes neighboring the boundary, and $ \bf{n} $ is the boundary normal vector. The boundary temperature can be calculated as,
\begin{eqnarray}\label{e28}
{T_w} = \frac{{9{b_1}{T_1} - {b_1}{T_2} + 3{\bf{n}} \cdot {{\bf{e}}_i}\Delta x{b_3}}}{{8{b_1} + 3{\bf{n}} \cdot {{\bf{e}}_i}\Delta x{b_2}}}.
\end{eqnarray}
After obtaining ${T_w}$, the unknown distribution functions $ {g_{\vec i}}({{\bf{x}}_f},t + \Delta t) $ can be calculated using Eq. (\ref{e26}).

\section{Numerical experiments}\label{sec.3}
In this section, several benchmark problems are conducted to verify our implementation of the heat source  and 
boundary conditions. In the present CLBM for the temperature field, the value of ${c_{T}}$ can be independent of ${c_s}$, and is set to be ${c_{T}} = \sqrt {2/5}$ in this work. Unless otherwise specified, the half-way bounce-back boundary scheme is used for both velocity and temperature boundary conditions, while $ {s_3} $ in Eq. (\ref{e8}) is chosen according to the non-slip rule ${s_3} = (16 - 8{s_\nu})/(8 - {s_\nu})$ \cite{fei2017consistent}.
\subsection{Time-independent diffusion problem}\label{sec.3.1}
\begin{figure*}[!ht]
	\center {
		{\epsfig{file=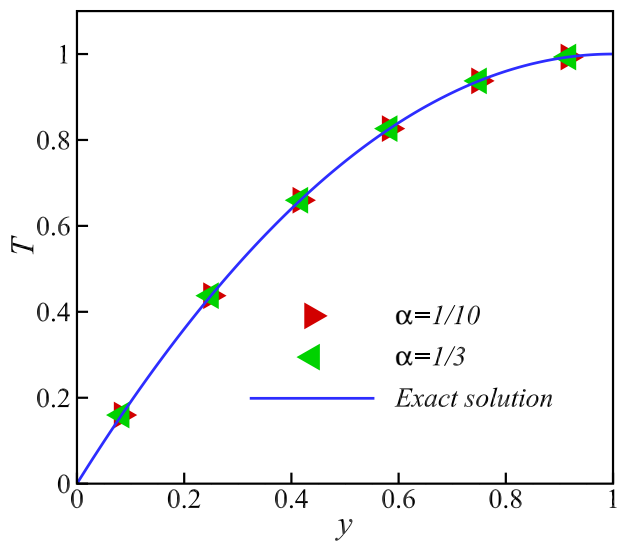,width=0.5\textwidth,clip=}}	
	}
	\caption{Comparison of temperature profiles predicted by the D2Q5 CLBM simulation and the analytical solution.}
	\label{FIG1}
\end{figure*}

\begin{figure*}[!ht]
	\center {
		{\epsfig{file=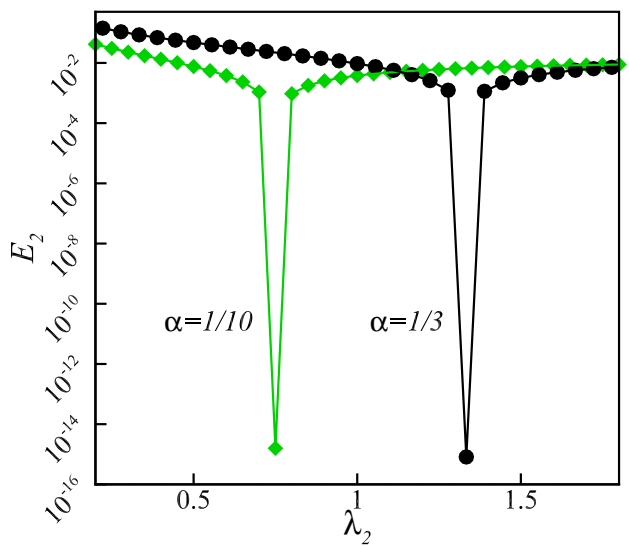,width=0.5\textwidth,clip=}}	
	}
	\caption{Global relative errors ${E_2}$ change with $ {\lambda _2} $ for $\alpha=1/10$ and $\alpha=1/3$.}
	\label{FIG2}
\end{figure*}

The first tested problem is a time-independent diffusion problem, which can be described by the following simplified equation and boundary conditions,
\begin{subequations}\label{e29}
	\begin{equation}
	\alpha \frac{{{\partial ^2}T}}{{\partial {y^2}}} + Q = 0,
	\end{equation} 
	\begin{equation}
	T(x,y = 0) = {T_0}, ~~{\rm{  }}T(x,y = L) = {T_L},
	\end{equation} 
\end{subequations}
where $ {T_0} $ and $ {T_L} $ are the temperatures at the bottom and the top of a straight channel. The heat source is $Q = 2\alpha \Delta T/{L^2}$, with $ \Delta T = ({T_L} - {T_0})$, and the exact solution is, 
\begin{equation}
T_a = {T_0} + \frac{{\Delta Ty}}{L}(2 - \frac{y}{L}).
\end{equation}
Due to the simple flow configuration, only 6 nodes are used to cover the channel width ($L = 6\Delta x$).  The simulation results are compared with the analytical solution in Fig. \ref{FIG1}. Two cases with $\alpha  = \left[ {1/10,1/3} \right]$ are considered, where the boundary conditions are ${T_0} = 0$ and ${T_L} = 1$, respectively. The corresponding relaxation rates are chosen as: (1)
${\lambda _1} = 4/3$ and ${\lambda _2} = 3/4$ for the first case;  (2) ${\lambda _1} = 3/4$ and ${\lambda _2} = 4/3$ for the second case. It is seen that the simulation results are in very good agreement with the analytical solution.
As analyzed by Cui \textit{et al.} \cite{cui2016discrete}, when the relaxation rate $ {\lambda _2} $ is specified as ${\lambda _2} = 12({\lambda _1} - 2)/({\lambda _1} - 12)$, the numerical slip in the D2Q5 MRT can be eliminated. To check its applicability in the present D2Q5 CLBM, a series of simulations are carried out with $ {\lambda _2} $ changing from 0.2 to 1.8. As shown in Fig. \ref{FIG2}, the global relative error ${E_2}$, defined as ${E_2}{\rm{ = }}\sqrt {\sum {{(T - {T_a})}^2}/\sum {T_a}^2} $, reaches the minimum
values at ${\lambda _2} = 3/4$ and ${\lambda _2} = 4/3$ for 
$\alpha=1/10$ and $\alpha=1/3$, respectively. Thus the non-slip rule in the D2Q5 MRT is also suitable for the present D2Q5 CLBM, which further verifies our previous analysis that the MRT-LBM and CLBM can be put into a unified general framework \cite{fei2017consistent}.
\subsection{Viscous dissipation in Poiseuille flow}\label{sec.3.2}
To validate the implementation of a spatially variable heat source, viscous dissipation in Poiseuille flow is simulated. The flow is driven by a constant body force along $ x $ direction, ${\bf{F}} = [{F_x},0]$, while the walls are at constant temperature $ T_w $. The viscous dissipation is considered by adding a heat source, $Q = \nu {(\partial u/\partial y)^2}$, in Eq. (\ref{e1c}). By using the non-slip rule for the velocity field, a very accurate velocity profile can be provided by the D2Q9 CLBM in  Section. \ref{sec.2.1}. The analytical temperature field is \cite{wang2007lattice},
\begin{equation}
T_a = {T_w} + \frac{1}{{3\nu \alpha }}{\left( {\frac{{{h^2}{F_x}}}{2}} \right)^2} \left[ {1 - {{\left( {\frac{y}{h}} \right)}^4}} \right],
\end{equation} 
where $ h $ is the half-width of the channel.
\begin{figure*}[!ht]
	\center {
		{\epsfig{file=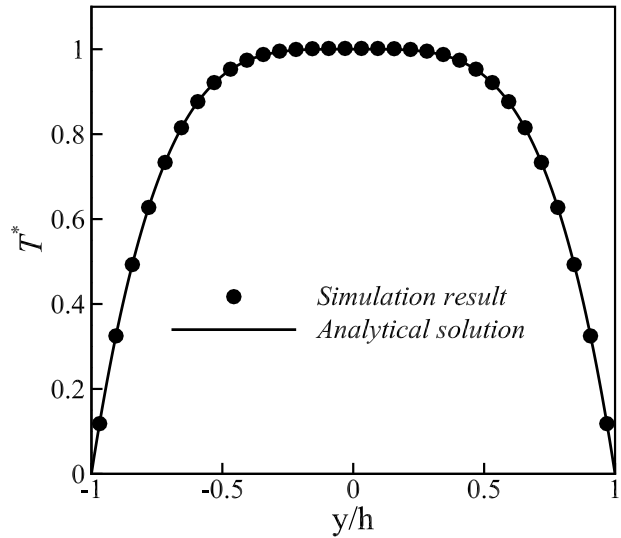,width=0.5\textwidth,clip=}}	
	}
	\caption{Comparison of the dimensionless temperature profiles predicted by the D2Q5 CLBM simulation and the analytical solution.}
	\label{FIG3}
\end{figure*}
The simulation result is compared with the analytical solution in Fig. \ref{FIG3}, where
the dimensionless temperature is defined as,
\begin{eqnarray}
{T^*} = 3\nu \alpha (T - {T_w})/{\left( {\frac{{{h^2}{F_x}}}{2}} \right)^2}.
\end{eqnarray}
It is clearly shown that the simulation result agrees well with the analytical solution.
\begin{figure*}[!ht]
	\center {
		{\epsfig{file=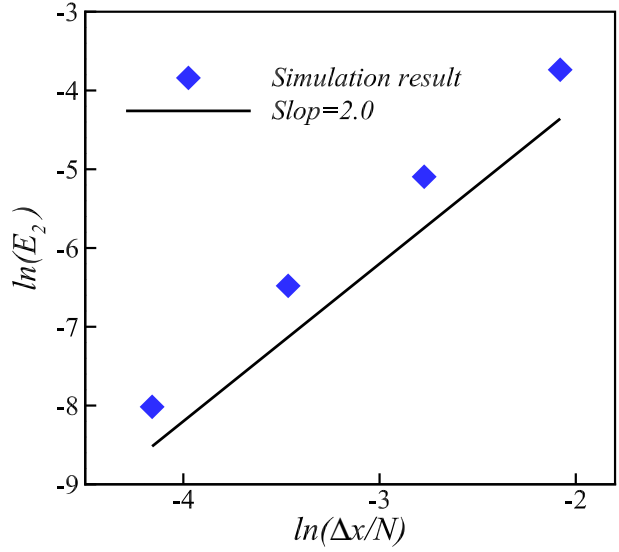,width=0.5\textwidth,clip=}}	
	}
	\caption{Global relative errors ${E_2}$ change with grid sizes for viscous dissipation problem in Poiseuille flow.}
	\label{FIG4}
\end{figure*}
The global relative errors at different grid sizes are shown in Fig. \ref{FIG4}. A very good 
linear fit is seen in the simulation results, and the slop is 2.05. It indicates that the implementation of the boundary conditions and heat source for the present problem has second-order accuracy in space.
\subsection{Natural convection in a square cavity}\label{sec.3.3}
The natural convection driven by the buoyancy force in a square cavity is simulated to validate the implementation of complex thermal boundary conditions. This problem has been widely examined in the literatures \cite{fei2016thermal,shu1998comparison,guo2002coupled,peng2003simplified}. The left and right walls of a square cavity are at constant temperature ${T_h} = 1$ and ${T_l} = 0$, respectively, while the top and the bottom walls are adiabatic. The problem is characterized by the Prandtl number ${{Pr}} = \nu /a$ and Rayleigh number $Ra = g\beta ({T_h} - {T_l}){H^3}/(\nu a)$, where $ H $ is the cavity hight. In the present paper, $ Pr $ is set to be 0.71, and the characteristic velocity $U = \sqrt {g\beta ({T_h} - {T_l})H}$ is set to be 0.1. The grid sizes are chosen to be $Nx \times Ny = 128 \times 128$, $ 192 \times 192 $, $  192 \times 192 $ and $  256 \times 256 $ for $Ra = {10^3}$, $ {10^4} $, $ {10^5} $ and $ {10^6} $, respectively.
\begin{figure*}[!ht]
	\center {
		\subfigure[]
		{\epsfig{file=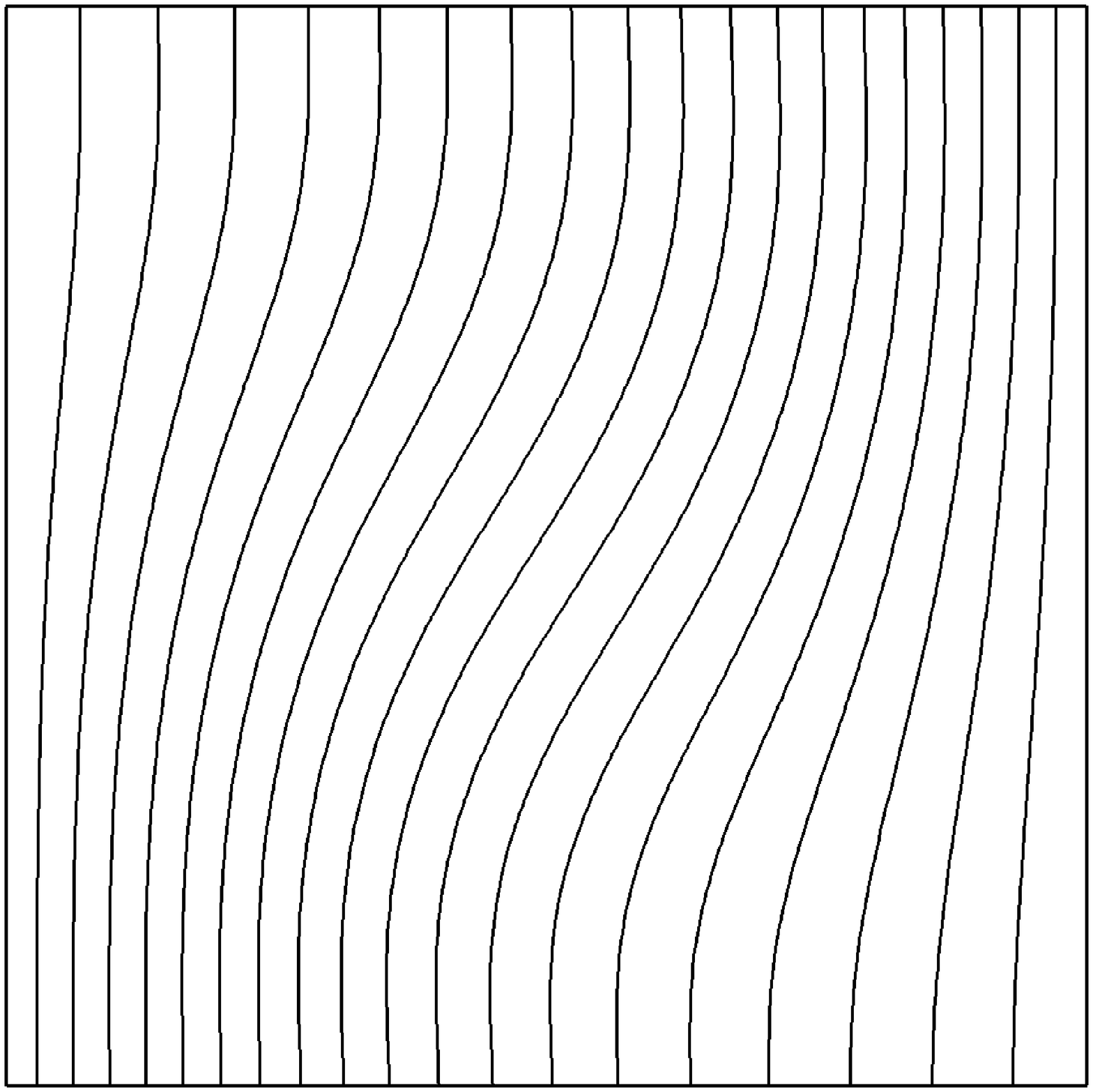,angle=-90,width=0.45\textwidth,clip=}}\hspace{0.5cm}
		\subfigure[] 
		{\epsfig{file=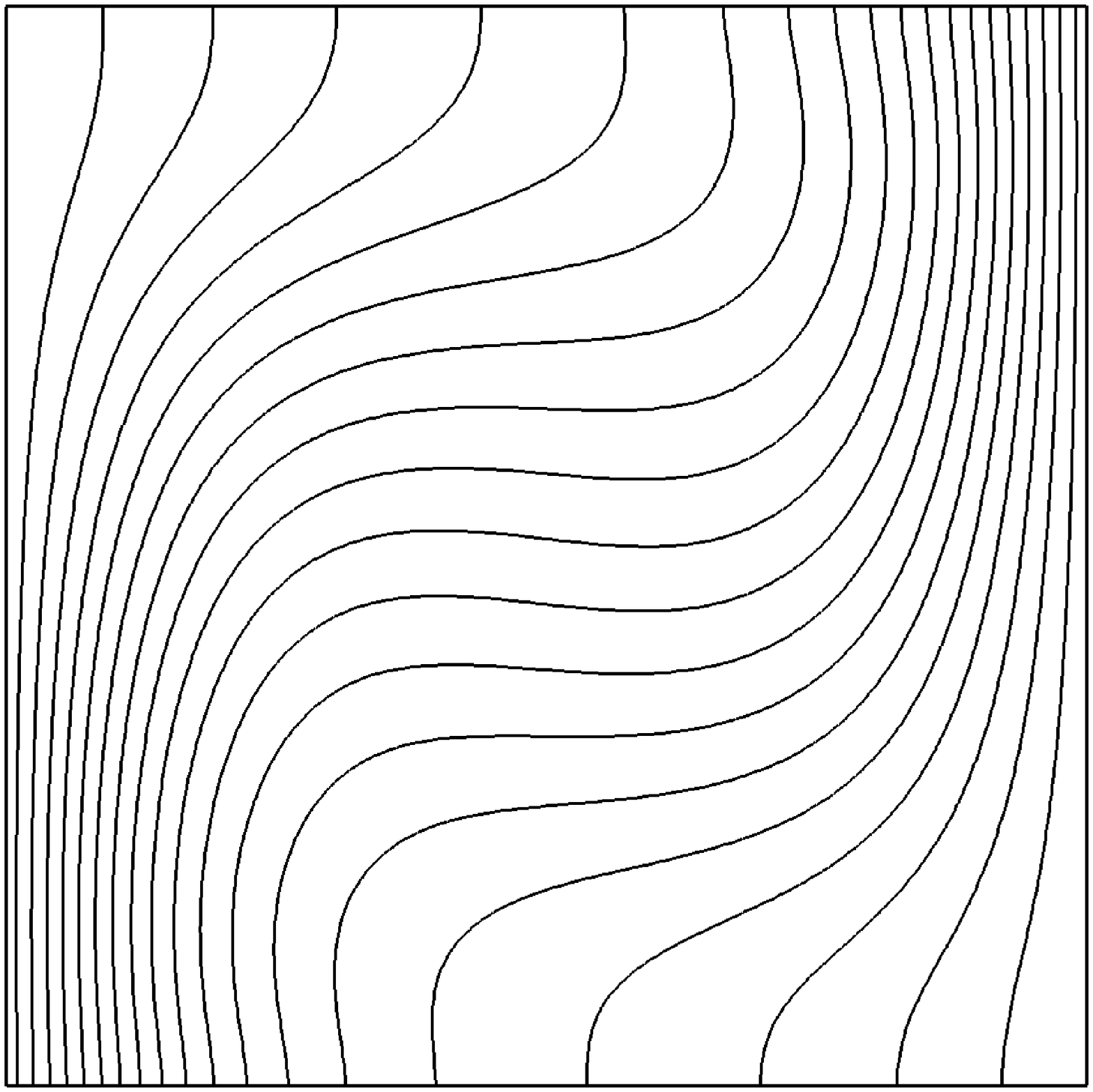,angle=-90,width=0.45\textwidth,clip=}}\vspace{0.0cm}\\
		\subfigure[]
		{\epsfig{file=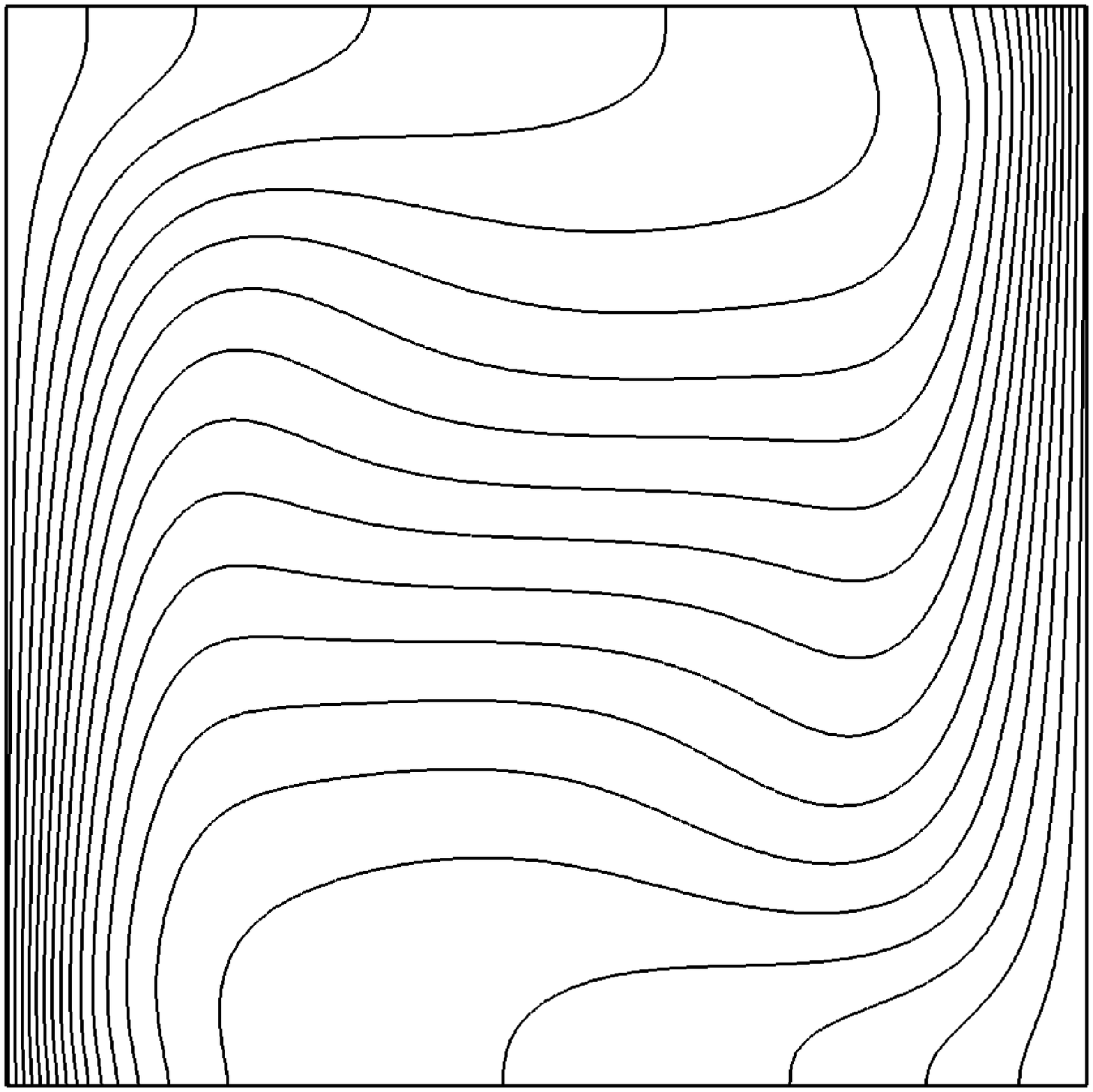,angle=-90,width=0.45\textwidth,clip=}}\hspace{0.5cm}
		\subfigure[]  
		{\epsfig{file=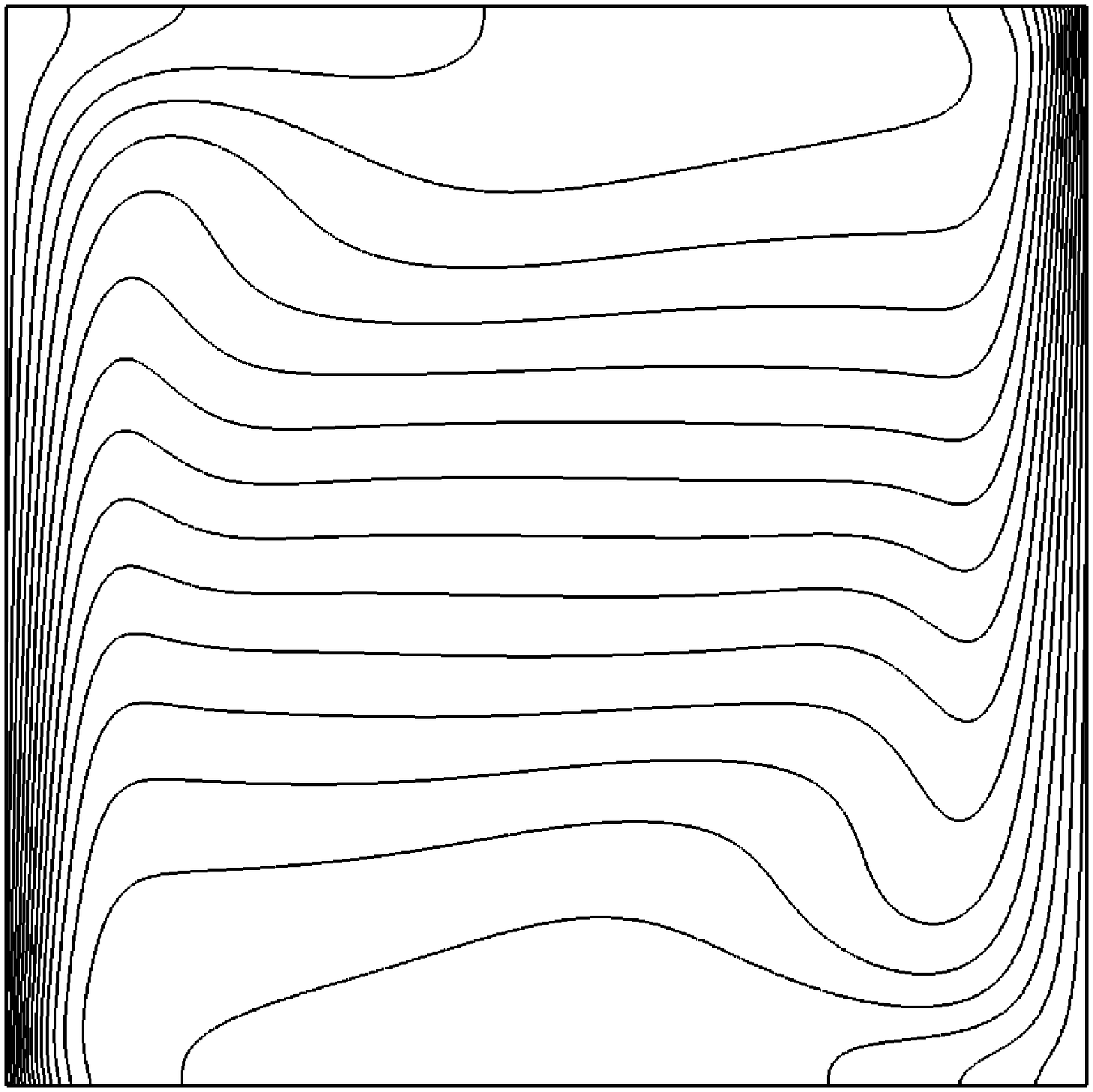,angle=-90,width=0.45\textwidth,clip=}}\vspace{0.0cm}\\
	}
	\caption{Isotherms of natural convection in a square cavity at: (a) $ Ra = {10^3}$, (b) $ Ra = {10^4} $, (c) $ Ra = {10^5} $, and (d) $ Ra = {10^6} $.}
	\label{FIG5}
\end{figure*}
The isotherms and streamlines at different $ Ra $ are shown in Figs. \ref{FIG5} and \ref{FIG6}, respectively. Qualitatively, all the characteristics in both temperature and flow fields agree well with the results in previous studies \cite{shu1998comparison,guo2002coupled,peng2003simplified}. To be more quantitative, data of the present work are listed in Table \ref{TAB1}, compared with those reported in previous studies \cite{peng2003simplified,guo2002coupled}. The following quantities are compared: the maximum horizontal velocity component ${u_{\max }}$ at $x = H/2$ and its location ${y_{\max }}$, the maximum vertical velocity component ${v_{\max }}$ at $y = H/2$ and its location ${x_{\max }}$, and the average Nusselt number $Nu$ along the cold wall. There is an excellent agreement between the present results and the benchmark solutions in the previous studies \cite{guo2002coupled,peng2003simplified}.
\begin{figure*}[!ht]
	\center {
		\subfigure[]
		{\epsfig{file=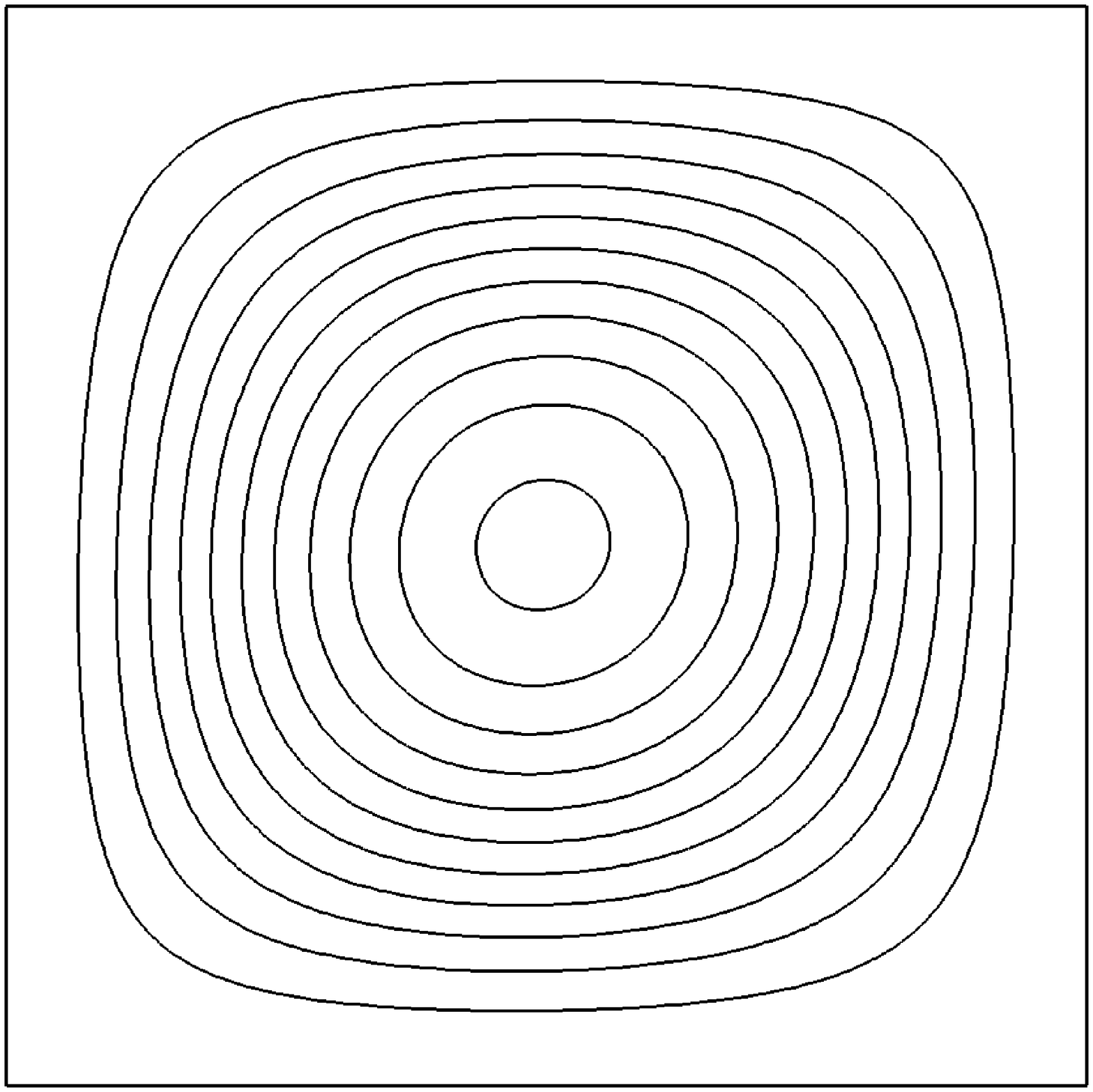,angle=-90,width=0.45\textwidth,clip=}}\hspace{0.5cm}
		\subfigure[] 
		{\epsfig{file=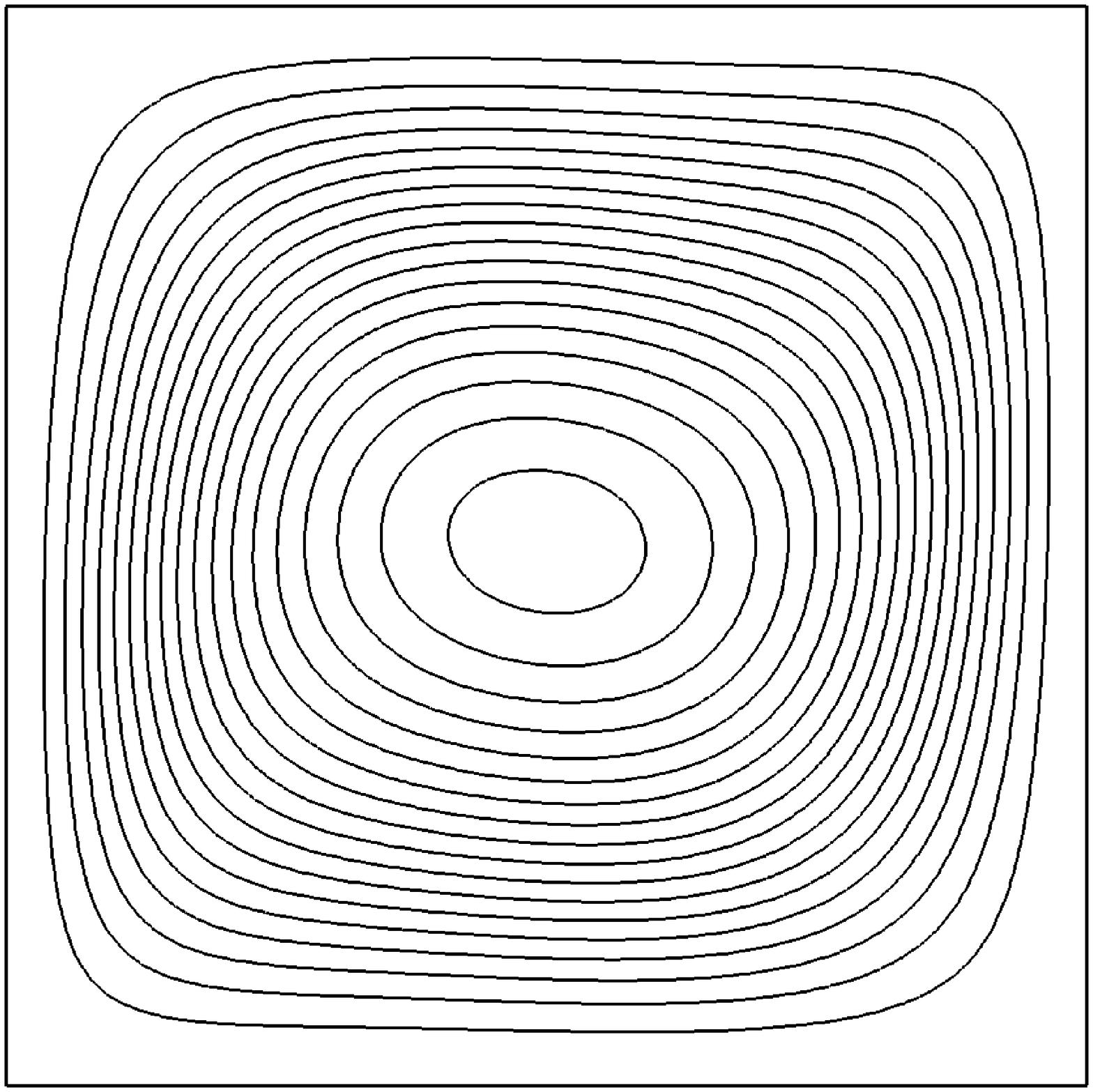,angle=-90,width=0.45\textwidth,clip=}}\vspace{0.0cm}\\
		\subfigure[]
		{\epsfig{file=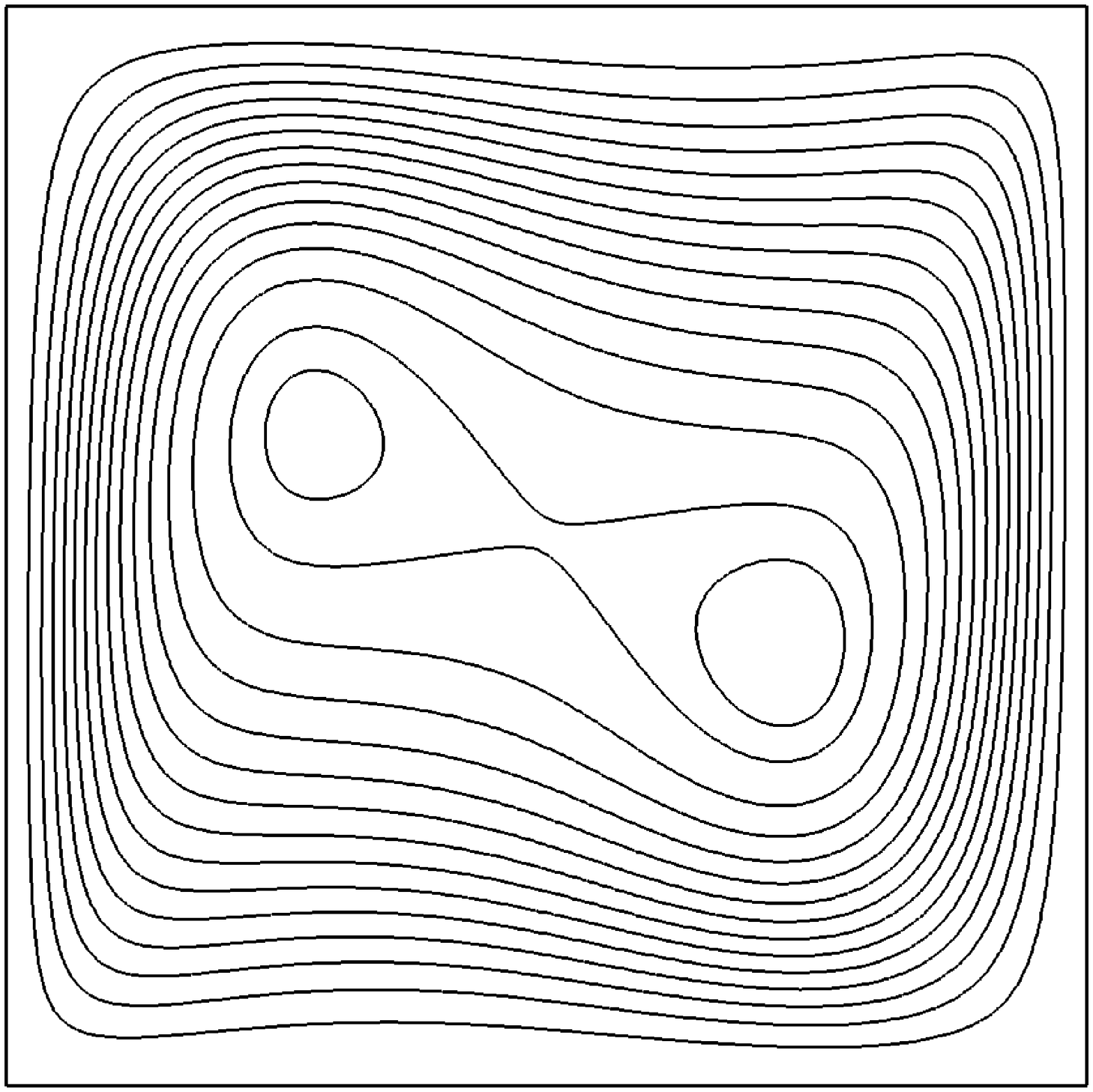,angle=-90,width=0.45\textwidth,clip=}}\hspace{0.5cm}
		\subfigure[]  
		{\epsfig{file=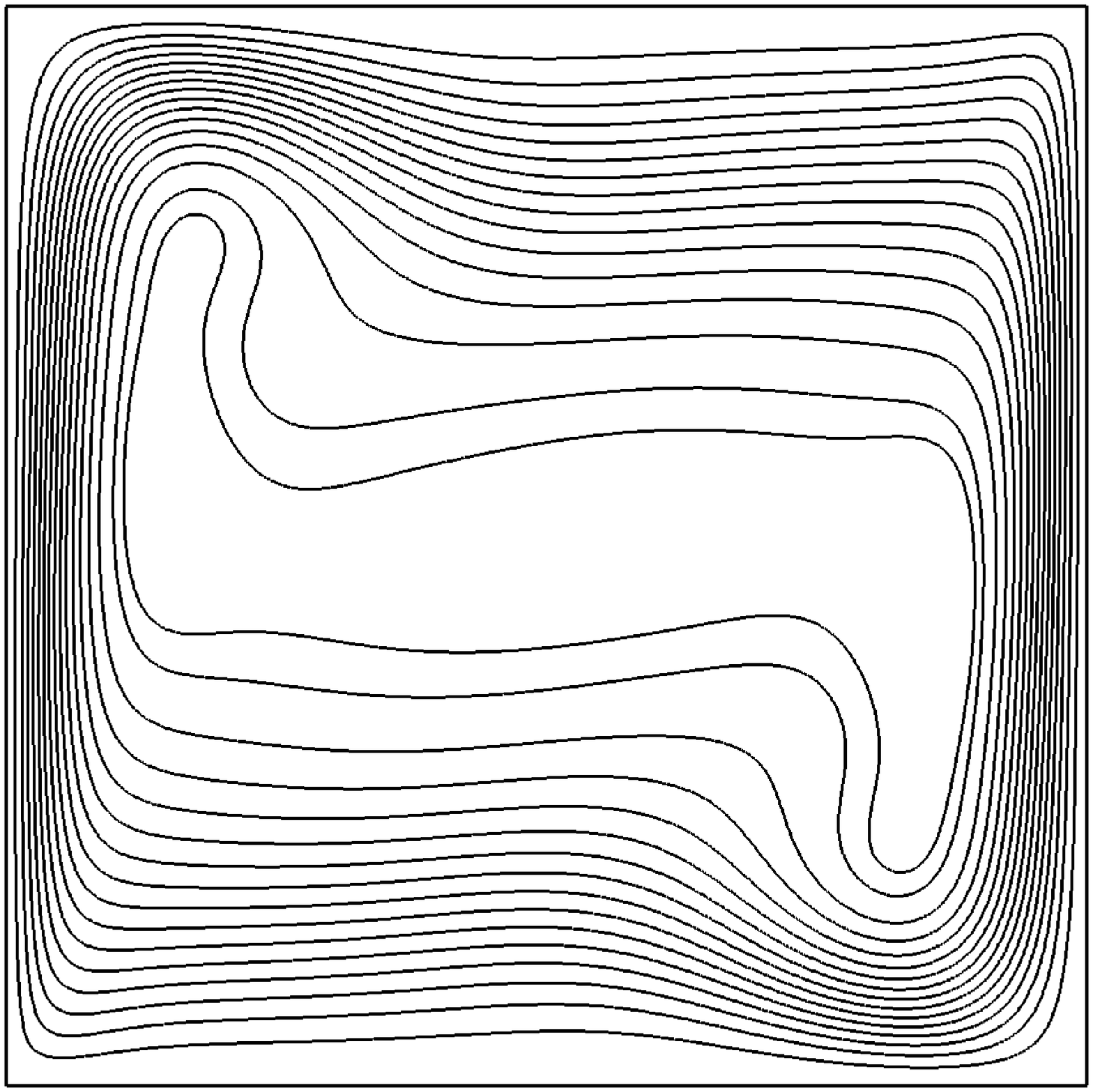,angle=-90,width=0.45\textwidth,clip=}}\vspace{0.0cm}\\
	}
	\caption{Streamlines of natural convection in a square cavity at: (a) $ Ra = {10^3}$, (b) $ Ra = {10^4} $, (c) $ Ra = {10^5} $, and (d) $ Ra = {10^6} $.}
	\label{FIG6}
\end{figure*}

\begin{table}[htbp]
	\caption{Comparisons of the present CLBM results with the Benchmark solutions \cite{peng2003simplified,guo2002coupled}.}
	\centering
	\label{TAB1}
	\begin{tabular}{ l   l   l   l   l   l   }
		\toprule
		$ Ra $&	 &  $ {10^3} $ & $ {10^4}
		$ & $ {10^5}$ & $ {10^6}$\\ \hline
		$ {{u_{\max }}} $ &LBM1 \cite{peng2003simplified}& 3.644 &  16.134  & 34.87 & 64.838 \\ 
		$  $	 &LBM2 \cite{guo2002coupled}& 3.6554 &  16.0761
		& 34.8343 & 65.3606 \\
		$  $	&CLBM& 3.6532 &  16.1737  & 35.0488 & 65.0274 \\
		
		${y_{\max }}$ &LBM1 \cite{peng2003simplified}& 0.815 &  0.825  & 0.855 & 0.850 \\ 
		$  $	 &LBM2 \cite{guo2002coupled}& 0.8125 &  0.8203
		& 0.8594 & 0.8516 \\
		$  $	&CLBM& 0.8140 &  0.8255  & 0.8574 & 0.8525 \\
		
		${{v_{\max }}}$ &LBM1 \cite{peng2003simplified}& 3.691 &  19.552  & 67.799 & 215.26 \\ 
		$  $	 &LBM2 \cite{guo2002coupled}& 3.6985 &  19.6368
		& 68.2671 & 216.415 \\
		$  $	&CLBM& 3.6999 &  19.6735  & 68.7584& 220.919 \\
		
		${x_{\max }}$ &LBM1 \cite{peng2003simplified}& 0.180 &  0.120  & 0.065 & 0.040 \\ 
		$  $	 &LBM2 \cite{guo2002coupled}& 0.1797 &  0.1172  & 0.0625 & 0.0391  \\
		$  $	&CLBM& 0.1792 &  0.1172  & 0.0647 & 0.0387\\ 
		
		$Nu$ &LBM1 \cite{peng2003simplified}& 1.117 &  2.241  & 4.491 & 8.731 \\ 
		$  $	 &LBM2 \cite{guo2002coupled}& 1.1168 &  2.2477  & 4.5345 & 8.7775  \\
		$  $	&CLBM& 1.1174 &  2.2428  & 4.5178 & 8.8204\\ 
		
		\toprule  
	\end{tabular}
\end{table}  
\section{Conclusions}\label{sec.4}
In this work, we extend previous DDF-based thermal CLBM to simulate more general incompressible thermal flows with heat sources and thermal boundary conditions. To include a heat source in the temperature equation,
a discrete source term ${R_i}$ is added to the collision step in central-moment space. To 
deal with thermal boundary conditions, the general bounce-back boundary scheme in MRT-LBM is modified and adopted in the present D2Q5 CLBM. Through numerical simulations  of several benchmark cases, very good accuracy of the proposed implementation for the heat source and boundary conditions are confirmed. In addition, it is found that the non-slip rule in the D2Q5 MRT-LBM is also suitable for the D2Q5 CLBM.
\section *{Acknowledgments}
Support from the MOST National Key Research and Development Programme (Project No. 2016YFB0600805) and the Center for Combustion Energy at Tsinghua University is gratefully acknowledged. The simulations were partly performed on the Tsinghua High-Performance Parallel Computer supported by the Tsinghua National Laboratory for Information Science and Technology and partly on ARCHER funded under the EPSRC project ``UK Consortium on Mesoscale Engineering Sciences (UKCOMES)" (Grant No. EP/L00030X/1). 

\section*{References}

\bibliography{DSFD2017}
\end{document}